\documentclass[12pt,preprint]{aastex}
\usepackage{amsmath}

\shorttitle{Towards Synchronism Through Dynamic Tides in J0651: the ``Antiresonance'' Locking}
\shortauthors{Valsecchi et al.}

\begin{document}


\title{Towards Synchronism Through Dynamic Tides in J0651: the ``Antiresonance'' Locking}


\author{F. Valsecchi\altaffilmark{1}, W. M. Farr\altaffilmark{1},
  B. Willems\altaffilmark{1}, and V. Kalogera\altaffilmark{1}}
\affil{Center for Interdisciplinary Exploration and Research in
  Astrophysics (CIERA) and Department of Physics and Astronomy,
  Northwestern University, 2145 Sheridan Road, Evanston, IL 60208,
  USA.}  \email{francesca@u.northwestern.edu}


\begin{abstract}
In recent years, the Extremely Low Mass White Dwarf (ELM WD) survey
has quintupled the number of known close, detached double WD binaries
(DWD). The tightest such DWD, SDSS J065133.33+284423.3 (J0651),
harbors a He WD eclipsing a C/O WD every $\simeq\,12\,$min. The
orbital decay of this source was recently measured to be consistent
with general relativistic (GR) radiation. 
Here we investigate the role of dynamic tides in a J0651-Like binary
and we uncover the potentially new phenomenon of ``antiresonance'' locking.
In the most probable scenario of an asynchronous binary at birth, we find that
 dynamic tides play a key role in explaining the measured GR-driven
  orbital decay, as they lock the system at stable 
antiresonances with the star's eigenfrequencies.
We show how such locking is naturally achieved and how, while locked at 
an antiresonance, GR drives the evolution of the orbital separation, 
while dynamic tides act to synchronize the spin of the He WD with the companion's orbital
motion, but \emph{only on the GR timescale}.  Given the relevant orbital and spin evolution timescales, 
the system is clearly on its way to synchronism, if not already synchronized.
\end{abstract}


\section{Introduction} \label{Intro}
Short period DWDs are the most numerous sources for the
next generation of low-frequency (mHz) space-based gravitational wave
(GW) detectors (e.g. LISA, \citealt{Danzmann1996}, \citealt{Hughes2006},
and eLISA/NGO, \citealt{NGOeLISA2012}).  The eclipsing DWD SDSS
J065133.33+284423.3 (hereafter J0651, \citealt{BrownEtAl2011,
  HermesEtAl2012}), discovered by the ELM WD survey
\citep{BrownEtAl2010, BrownEtAl2012, KilicEtAl2010, KilicEtAl2011},
harbors a tidally deformed He WD orbiting a C/O WD every $\simeq
12\,$min; this is the shortest-period detached system known. Here, we
investigate the effects of dynamic tides on the orbital evolution of
this binary.

Previous investigations of tides in J0651 placed limit on
 the components' tidal parameter $Q$, assuming tidal heating
 to be the cause of the observed luminosity \citep{Piro2011}, or
 used polytropes to estimate the deviation from the pure GR-driven inspiral, 
 under the assumption that the components are synchronized \citep{Benacquista2011}.

The orbital evolution of detached DWDs is often assumed to be driven
by GR emission alone.  
\cite{WDK2010} validated this assumption provided that the binary is in the regime of
 dissipative quasi-static tides 
(in the limit of orbital and rotational periods $\gg \sqrt{R^{3}(GM)^{-1}}$). 
The inefficiency of quasi-static tides means GR emission eventually 
drives the system out of synchronism and DWDs enter a regime in which dynamic
tides and resonantly excited gravity modes ($g$-modes) must be taken
into account.

Here, we use our newly developed
tidally excited stellar pulsation code {\it CAFein} (Valsecchi et
al.\,to be submitted) and detailed models of a J0651-Like He WD
 to study in detail the effect of radiative
dissipation of dynamic tides on the orbital evolution of J0651.
Our focus on radiative dissipation follows our understanding of 
dissipative dynamic tides in non-degenerate stars with radiative envelopes 
\citep{Zahn1975, WitteSavonije2002TwoRitatingMSstars}.

\cite{FullerLai2011a, FullerLai2012Dynamic} investigated the impact of 
\textit{resonantly} excited $g-$modes on the orbital evolution of  C/O DWDs. 
Dissipation in \cite{FullerLai2012Dynamic} was treated via the so-called outgoing 
wave boundary condition (BC), which implicitly assumes that tides 
are efficiently damped via radiative damping or non-linear effects. 
The authors focused on the evolution of $g-$mode amplitudes  
{\it during} resonance, without a detailed treatment of tidal dissipation 
in and \textit{out} of resonance, and found that resonant mode excitation could 
quickly drive the binary to (near) synchronization 
of the spin and orbit on the dynamic-tides timescale. 
Here, we reach fundamentally different conclusions: we find that, for a broad range of 
initial spin frequencies, dynamic tides act to drive the
spin-orbit frequency mismatch to and trap it in an
\emph{anti}resonance, where the dissipation due to tides effectively
vanishes. The orbit subsequently evolves due to GR emission, and the
only action of tides is to keep the spin-orbit frequency difference
trapped at the antiresonance value; as GR emission proceeds, 
both the spin and orbital frequencies increase, while their difference is locked, 
and the system is driven to synchronism, but this
process takes place on the GR timescale.

Very recently, \cite{HermesEtAl2012} reported the detection of orbital
decay in J0651 consistent with the GR prediction. 
Here we show that dynamic tides potentially play a key role in
such measurement, as they naturally drive the system into
\textit{antiresonance locking}, a new phenomenon we identify, with further
evolution of the orbit occurring only on the GR timescale.

\section{Observed and Model's Properties} \label{sec:Observations and Model's Properties}
The DWD system J0651 was originally discovered by \citet{BrownEtAl2011}.
\citet{HermesEtAl2012} measured an effective temperature, gravity, radius, and mass for the He WD of 
$T_{\rm eff} =16530\,\pm\,200\,$K, ${\rm
  log}(g)\,=\,6.76\,\pm\,0.04\,$dex, $R_{\rm
  1}\,=\,0.0371\,\pm\,0.0012\,R_{\odot}$, and $M_{\rm
  1}\,=\,0.26\,\pm\,0.04\, M_{\odot}$; 
  they also measured the C/O WD mass to be $0.5\,\pm\,0.04\,M_{\odot}$ 
  and the orbit is circular. 
  The measured orbital period and orbital decay are $P\,=\,765.206543(55)\,$s and
$\dot{P}\,=\,(-9.8\,\pm\,2.8)\times\,10^{-12}$s$\,$s$^{-1}$, respectively. For comparison, the
orbital period decay expected from GR emission alone is
$\dot{P}\,=\,(-8.2\,\pm\,1.7)\times\,10^{-12}$s$\,$s$^{-1}$ for the masses derived
by \cite{HermesEtAl2012}.  We use MESA \citep{PBDHLT2011} to create a
He WD model with $T_{\rm eff}\,\simeq\,16766\,$K, ${\rm
  log}(g)\,=\,6.68\,$dex, $R_{\rm 1}\,=\, 0.0363\,R_{\odot}$, and
$M_{\rm 1}\,=\,0.23\, M_{\odot}$.  These parameters are within
$1\,\sigma$ of the measured $M_{\rm 1}$ and $R_{\rm 1}$ and
$2\,\sigma$ of the measured $T_{\rm eff}$ and log$(g)$.

\section{Tidal Evolution} \label{sec:Main Equations}
\subsection{Non-Adiabatic Stellar and Tidal Pulsations} \label{sec:Non-adiabatic stellar pulsation equations and dynamic tides}
When tides are dynamic, the tidal forcing exerted by a star in a
binary can interact with one or more of the companion's
eigenfrequencies.  Here we take the He WD (primary) to
rotate uniformly with an angular velocity $\Omega_{\rm 1}$ and, given
the MESA model's current cooling timescale $t_{\rm cool}$  (see below), we
assume that its stellar parameters do not evolve.  
We also neglect the
effects of Coriolis and centrifugal forces (the break-up spin for the model is $\simeq
2.5\,$min).  Under these assumptions,
the star's eigenmodes correspond to those of a non-rotating
spherically symmetric star in hydrostatic equilibrium. We treat the
C/O WD (secondary) as a point mass.
Let ${\bf r} = (r, \theta, \phi)$ be a system of spherical coordinates
with respect to an orthogonal frame co-rotating with the primary. The
system of equations governing non-adiabatic and non-radial stellar
pulsations has been derived e.g. by \cite{UnnoEtAl1989} (Eqs. 24.7 -
24.12 in \S 24).  It comprises six homogeneous ordinary differential
equations involving the complex dimensionless eigenfrequency $\omega$
($\sigma=\omega \sqrt{G M / R^3}$, where $\sigma$ is the dimensional
eigenfrequency).  It is convenient to use
\begin{align}
&y_{\rm 1} = \frac{\xi_{\rm r}}{r},\,\,\,y_{\rm 2} =
  \frac{1}{gr}\left(\frac{p'}{\rho}+ \Phi'\right) = \frac{\omega^2
    r}{g} \frac{\xi_{\rm h}}{r},\,\,\,y_{\rm 3} =
  \frac{1}{gr}\Phi'\nonumber\\ &y_{\rm 4} =
  \frac{1}{g}\frac{d\Phi'}{dr},\,\,\,y_{\rm 5} = \frac{\delta
    S}{c_{\rm p}},\,\,\,y_{\rm 6} = \frac{\delta L_{\rm R}}{L_{\rm R}}
\end{align}
as integration variables in the eigenvalue problem.  Here $\xi_{\rm
  r}$ and $\xi_{\rm h}$ are the radial and orthogonal component of the
displacement of the star's mass element from the equilibrium
position, $\rho$ the density, $c_{\rm p}$ the specific heat at
constant pressure, and $p', \Phi', \delta S$, and $\delta L_{\rm R}$
the perturbed pressure, potential of self-gravity,
entropy and radiative luminosity, respectively. The effect of
convection on the oscillations is neglected.

The central BCs are that the entropy is constant during
a single oscillation ($\delta S\,=\,0$) and that $\Phi'$,
$(p'/\rho+\Phi')$, and $\xi_{\rm r}$ are regular.  The
surface BCs are determined by assuming that near the surface the
pressure and the density drop steeply outward, by considering that
there is no inward radiative flux, and by requiring the continuity of
$\Phi'$ and its first derivative. 

The tidal action from the companion is included as a small
time-dependent force which perturbs the hydrostatic equilibrium of a
spherically symmetric static star. The system of equations describing
the tidal response of the star is formally identical to the equations
describing pure stellar pulsations, provided that the perturbed
potential of self-gravity $\Phi'$ is replaced by the total
perturbation of the potential $\Psi\,=\,\Phi' + \epsilon_{\rm T}W$, 
where $\epsilon_{\rm T}W$ is the tide-generating potential. We expand
$\epsilon_{\rm T}W$ in a Fourier series as (see
e.g. \citealt{PolflietSmeyers1990})
\begin{equation}
\epsilon_{\rm T}W({\bf r}, t) = -\epsilon_{\rm T}\sum_{l=2}^{4}\sum_{m=-l}^{l}\sum_{k=-\infty}^{\infty}c_{l,m,k}\left(\frac{r}{R_{1}}\right)^{l} Y^m_l(\theta, \phi){\rm exp}[i(\sigma_{m,k}t-k\Omega_{\rm orb}\tau)]
\label{eq:tidalPotential}
\end{equation}
where $Y^m_l(\theta, \phi)$ are unnormalized spherical harmonics of
harmonic degree $l$, and azimuthal number $m$, and $k$ is the Fourier index;
$\theta$ is the colatitude in the WD's co-rotating frame, and
$\phi$ is the longitude.  At time $t=0$, the angle $\phi=0$ marks the
position of the periastron.  The dimensionless parameter
$\epsilon_{\rm T} \equiv (R_{\rm 1}/a)^3(M_{\rm 2}/M_{\rm 1})$ gives
the ratio of the tidal force to gravity at the star's equator.  The
semi-major axis of the binary is $a$, $\Omega_{\rm orb}$ is the mean
motion, $\tau$ the time at periastron passage, $\sigma_{m,k} =
k\Omega_{\rm orb}+m\Omega_{\rm 1}$ is the tidal forcing angular
frequency with respect to the co-rotating frame, and $c_{l,m,k}$ are
Fourier coefficients. For a binary with a circular orbit (like J0651)
the only non-zero $c_{l,m,k}$ are the ones with $k\,=\,-m$ and are defined as
\begin{equation}
c_{l,m,-m}  = \frac{(l-|m|)!}{(l+|m|)!}P^{|m|}_{l}(0)\left(\frac{R_{\rm 1}}{a}\right)^{l-2}
\label{eq:clmk_circular}
\end{equation}
where $P^{m}_{l}(x)$ are Legendre polynomials of the first kind. Refer
to \cite{PolflietSmeyers1990} for further discussion of $c_{l,m,k}$.
The expansion~(\ref{eq:tidalPotential}) shows that the tidal action
from the secondary induces in the primary an infinite number of
forcing angular frequencies $\sigma_{m,k}$. Tides are dynamic when
$\sigma_{m,k} \neq 0$.

In principle, all values of $l$, $m$, and $k$ must be considered in
the set of equations describing tidally excited stellar
pulsations. However, as $c_{l,m,k}\propto(R_{\rm 1}/a)^{l-2}$,
investigations on dynamic tides are often restricted to the terms
belonging to $l=2$. Here we consider only the leading term in the
expansion of $\epsilon_{\rm T}W$: ($l, m, k$) = (2,-2,2).  


\subsection{Secular Evolution of the Orbital Separation and Stellar Spin}\label{The Secular Evolution of the Orbital separation and stellar}
Here we summarize the main equations describing the
secular evolution of $a$ and $\Omega_{\rm 1}$ due to
dynamic tides for a binary with a circular orbit and for ($l, m, k$) = (2,-2,2); 
refer to \cite{WVHS2003,
  WDK2010} for a more detailed derivation and for an explanation of
the various parameters entering such equations.

The primary's tidal deformation perturbs the external gravitational
field, and thus the Keplerian motion of the binary components.  In the
framework of the theory of osculating elements in celestial mechanics
\citep{Sterne1960, BrouwerClemence1961, Fitzpatrick1970}, the rate of
change of the orbital semi-major axis due to the tidal deformation of
the star is given by
\begin{equation}
\left(\frac{da}{dt}\right)_{\rm sec} = \frac{8\pi}{P_{\rm orb}}\frac{M_{\rm 2}}{M_{\rm 1}}a \left(\frac{R_{\rm 1}}{a}\right)^{l+3} 
|F_{l,m,k}|{\rm sin}\gamma_{l,m,k}G^{(2)}_{l,m,k}(0)\label{eq:dadt_tides_final}
\end{equation}
where $G^{(2)}_{l,m,-m}(0) = -2mP^{|m|}_{l}(0)c_{l,m,-m}
$ for $e\,=\,0$, $P_{\rm orb}$ is the orbital period, and the dimensionless
$F_{l,m,k}$ are defined as
\begin{equation}
F_{l,m,k} = -\frac{1}{2}\left[\frac{R_{1}}{GM_{1}}\frac{\Psi_{l,m,k}(R_{\rm 1})}{\epsilon_{\rm T}c_{l,m,k}}+1\right]= |F_{l,m,k}|e^{i\gamma_{l,m,k}}
\label{eq:Flmk}
 \end{equation}
and measure the response of the star to the various tidal forcing
frequencies; $F_{l,m,k}$ can be computed from a solution to the
modified pulsation equations described above.  The angle
$\gamma_{l,m,k}$ describes the phase-lag between the tidal forcing and
the response of the total potential induced by the dissipation in the
system.  

This phase shift produces a torque on
the tidally deformed star, which acts on its spin yielding
\begin{equation}
\left(\frac{d\Omega_{\rm 1}}{dt}\right)_{\rm sec} =-
  \frac{8\pi}{P_{\rm orb}}\left(\frac{GM_{\rm 1}^{2}M_{\rm
      2}^{2}}{M_{\rm 1}+M_{\rm 2}}\right)^{1/2}\frac{M_{\rm 2}}{M_{\rm
      1}}\frac{a^{1/2}}{I_{\rm 1}}
	\left(\frac{R_{\rm
      1}}{a}\right)^{l+3}|F_{l,m,k}|{\rm
    sin}~\gamma_{l,m,k}\frac{G^{(2)}_{l,m,k}(0)}{2}
\label{eq:dOmegadt_tides_final}
\end{equation}
\section{The Dynamic tides regime in a J0651-Like binary} \label{sec:Dynamic tides in a J0651-like binary}
\begin{figure} [!h]
\epsscale{0.7}
\plotone{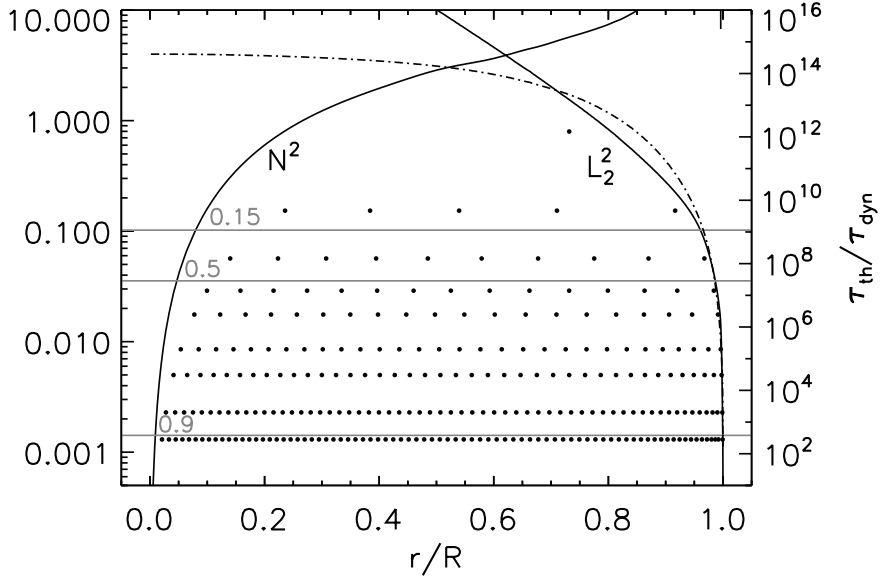}
\caption{{\it Left y-axis: } radial profile of the Br$\ddot{\rm
    u}$nt-Vaisala ($N$) and Lamb ($L_{\rm 2}$) frequencies squared (black-solid lines). The dots
  mark the positions of the nodes for the modes $g_{\rm 1}, g_{\rm 5},
  g_{\rm 10}, g_{\rm 15}, g_{\rm 20}, g_{\rm 30}, g_{\rm 40}, g_{\rm
    60}, g_{\rm 80}$. The horizontal grey lines mark the position of
  the tidal forcing frequencies $\omega_{\rm T}^{2}$ for spins (from
  top to bottom) $\Omega_{\rm 1}$ = 0.15 $\Omega_{\rm orb}$,
  $\Omega_{\rm 1}$ = 0.5 $\Omega_{\rm orb}$, and $\Omega_{\rm 1}$ =
  0.9 $\Omega_{\rm orb}$. All the frequencies are normalized by
  $\sqrt{GM/R^3}$. {\it Right y-axis: } the dashed-dot line shows the ratio of the star's thermal
   to dynamical ($\tau_{\rm dyn}$/$\tau_{\rm th}$) timescales. 
  Where $\tau_{\rm dyn}$/$\tau_{\rm th}$ is small non-adiabaticity becomes relevant.}
\label{fig:HeWD_0p23Msun_log1673_nodesAndPropagationZone_thermalToDynamical}
\end{figure}
\begin{table}[!h]
\caption{\label{Tab:l2_Eigenfrequencies_HeWD_0p23Msun_log1673}
  Non-adiabatic eigenfrequencies for a J0651-Like He WD. The
subscripts ``R'' and ``I'' denote the real and imaginary part,
respectively. Note that $\left |\omega_{\rm I}\right |$ increases with the order of
the mode, indicating that non-adiabaticity becomes relevant for
high-order modes 
(see Fig.~\ref{fig:HeWD_0p23Msun_log1673_nodesAndPropagationZone_thermalToDynamical}). 
Also, $\omega_{\rm I}<0$ for modes up to $g_{\rm 19}$, indicating that these modes are unstable.}
\begin{center}
\begin{tabular}{c|c|c}
\hline mode & $\omega_{\rm R}$ & $\omega_{\rm I}$ \\ \hline $g_{\rm
  1}$ & 8.928$\,\cdot\,10^{-1}$ & $\simeq\,10^{-14}$\\ $g_{\rm 5}$ &
3.915$\,\cdot\,10^{-1}$ & -1.65$\,\cdot\,10^{-11}$ \\ $g_{\rm 10}$ &
2.379$\,\cdot\,10^{-1}$ & -1.057$\,\cdot\,10^{-9}$ \\ $g_{\rm 15}$ &
1.701$\,\cdot\,10^{-1}$ & -7.804$\,\cdot\,10^{-9}$\\ $g_{\rm 20}$ &
1.326$\,\cdot\,10^{-1}$ & 8.403$\,\cdot\,10^{-9}$\\ $g_{\rm 30}$ &
9.237$\,\cdot\,10^{-2}$ & 1.959$\,\cdot\,10^{-6}$\\ $g_{\rm 40}$ &
7.065$\,\cdot\,10^{-2}$ & 9.882$\,\cdot\,10^{-6}$\\ $g_{\rm 60}$ &
4.785$\,\cdot\,10^{-2}$ & 2.842$\,\cdot\,10^{-5}$\\ $g_{\rm 80}$ &
3.611$\,\cdot\,10^{-2}$ & 5.519$\,\cdot\,10^{-5}$\\
\end{tabular}
\end{center}
\end{table}	
Here we first examine whether dynamic tides are relevant in a J0651-Like binary by
comparing the He WD eigenfrequencies with the tidal forcing
frequencies. We use {\it CAFein} (Valsecchi et al.\,to be submitted) to
compute the non-adiabatic $l\,=\,2$ eigenfrequencies for $g\,-\,$modes 
between $g_{\rm 1}$-$g_{\rm 80}$. The complex eigenfrequencies
are summarized in
Table~\ref{Tab:l2_Eigenfrequencies_HeWD_0p23Msun_log1673}. 
In
Fig.~\ref{fig:HeWD_0p23Msun_log1673_nodesAndPropagationZone_thermalToDynamical}
we show the propagation diagram for a J0651-Like He WD, where the
squared Br$\ddot{\rm u}$nt-Vaisala frequency is 
$N^2 = g(\Gamma_{\rm 1}^{-1}d{\rm ln}p/dr- d{\rm ln}\rho/dr)$ \citep{BrassardEtAl1991}, 
while the squared Lamb frequency is given by
$L_{\rm l}^2 = l(l+1)c_{\rm s}^2/r^2$.
The $g$-modes propagate in regions where $\omega_{\rm R}^{2}< N^2$ and
$L_{\rm l}^2$.  The dots in
Fig.~\ref{fig:HeWD_0p23Msun_log1673_nodesAndPropagationZone_thermalToDynamical}
mark the positions of the radial nodes for the modes considered in
Table~\ref{Tab:l2_Eigenfrequencies_HeWD_0p23Msun_log1673}, while the
horizontal grey lines mark the position of the tidal forcing frequency
($\omega_{\rm T}^{2}$) if $\Omega_{\rm 1}$ is varied.
As $\omega_{\rm T}^{2}$ and $\omega_{\rm R}^{2}$ cover the same frequency regime, 
resonances between tides and eigenmodes can be relevant
for a J0651-Like binary, and hence tides are dynamic in this system.
\section{Synchronism Through Dynamic Tides : the Antiresonance Locking} \label{sec:Synchronism Through Dynamic Tides : the Antiresonance Locking}
\begin{figure} [!h]
\epsscale{0.7}
\plotone{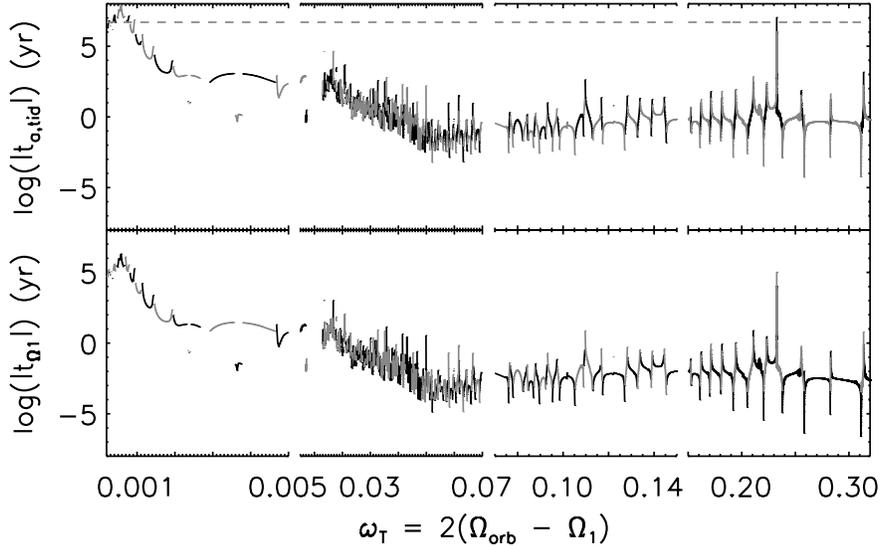}
\caption{Timescales for the secular evolution of orbital separation and stellar spin
due to dynamic tides. {\it Top: } ${\rm log}|t_{a, tid}| =
  {\rm log}|a/\dot{a}_{\rm sec}|$. The horizontal dashed line shows
  the theoretically expected GR timescale ${\rm log}|t_{\rm a, GR}|$ 
  from \cite{Peters1964} for the masses considered here
  ($t_{\rm a, GR}\,\simeq\,5\,$Myr). {\it Bottom: } ${\rm
    log}|t_{\Omega_{\rm 1}}| ={\rm log}|\Omega_{\rm
    1}/\dot{\Omega}_{\rm 1,sec}|$.  Grey and black lines and data
  points correspond to negative and positive timescales,
  respectively. As a reference, the first dip on the right marks the $g_{\rm 7}-$mode resonance.
Note that the x-axis is not uniform, and that
  $\left|t_{\Omega_1}\right| \simeq 10^{-2} \left| t_{\rm a,tid} \right|$.}
\label{fig:tidalTimescalesLog_vsAsynch_fullRange_broken}
\end{figure}
In Fig.~\ref{fig:tidalTimescalesLog_vsAsynch_fullRange_broken} we show
the timescales for the evolution of $a$ and $\Omega_{\rm 1}$ due to tides
calculated using Eqs.~(\ref{eq:dadt_tides_final}) and
(\ref{eq:dOmegadt_tides_final}).  We fix $\Omega_{\rm orb}$ and
$M_{\rm 2}$ to the currently observed values and vary $\Omega_{\rm 1}$,
thus changing $\omega_{\rm T}$. Here we consider $\Omega_{\rm 1}$
ranging between $\simeq\,$(0.15-0.999)$\,\Omega_{\rm orb}$.  Black and grey data points
represent positive and negative timescales, respectively. As expected,
$\dot{a}_{\rm sec} < 0$ when $\dot{\Omega}_{\rm 1, sec}
> 0$ and vice versa. 
Note that $\Omega_{\rm 1}$ evolves faster than $a$ by about 2 orders of magnitude . 
Dips in the timescales
correspond to resonances between the tidal forcing frequencies and the
star's eigenfrequencies, while the peaks mark the ``antiresonances''.

The sign of the timescales depends on the angle that
parametrizes tidal dissipation, $\gamma_{l,m,k}$. This angle, in turn,
is related to the physical angle $\phi$ describing the orientation of
the bulge in the star's reference frame via
\begin{equation}
\phi = -\gamma_{l,m,k}/m, 
\end{equation}
or for $m\,=\,$-2, $\phi\,=\,\gamma/2$.

When $\gamma_{2,-2,2} = 0$ or $\pi$, the rate of change of $a$ and
$\Omega_1$ becomes zero and the timescales display an antiresonance.  
At these lag angles the bulge is either aligned with the companion or $\pi/2$ out
of alignment; in either case there is no tidal torque and therefore no
change in the orbit or spin due to tides. On the other hand, resonances occur when
$\gamma_{2,-2,2}$ equals either $\pi/2$ or $-\pi/2$, and thus $\phi$
equal to $\pi/4$ or $-\pi/4$; in these configurations, the torque is
at a maximum.

The agreement between the observed orbital decay and the GR prediction 
implies that tides are not currently
affecting the orbit.  If the He WD were born close to synchronization
or evolved to synchronism with the companion's orbital motion, tides are
quasi-static ($\omega_{\rm T}\,\ll\,1$) or static ($\omega_{\rm
  T}\,=\,0$) and the orbital evolution timescales due to quasi-static
tides are longer than the theoretical GR timescale (see Fig.~\ref{fig:tidalTimescalesLog_vsAsynch_fullRange_broken}
or \citealt{WDK2010}).  On the other
hand, if the He WD were born in or evolved to an asynchronous state,
the signs of the timescales in
Fig.~\ref{fig:tidalTimescalesLog_vsAsynch_fullRange_broken} can result
in spin evolution that stably locks the star at antiresonances, where
 $t_{\rm a, tid}> t_{\rm a, GR}$ (see details below). The crossing of the
antiresonance peaks with the horizontal dashed line in
Fig.~\ref{fig:tidalTimescalesLog_vsAsynch_fullRange_broken} is visible
only for $\omega_{\rm T}\simeq 0.23$ because of the resolution adopted
during the scan of the parameter space in $\Omega_{\rm 1}$, but 
$t_\mathrm{a,tid} \gg t_\mathrm{a,GR}$ whenever the
angle $\gamma_{l,m,k}$ in Eq.~(\ref{eq:dadt_tides_final}) passes
though 0 and $\pi$.

\begin{figure} [!h]
\epsscale{0.7}
\plotone{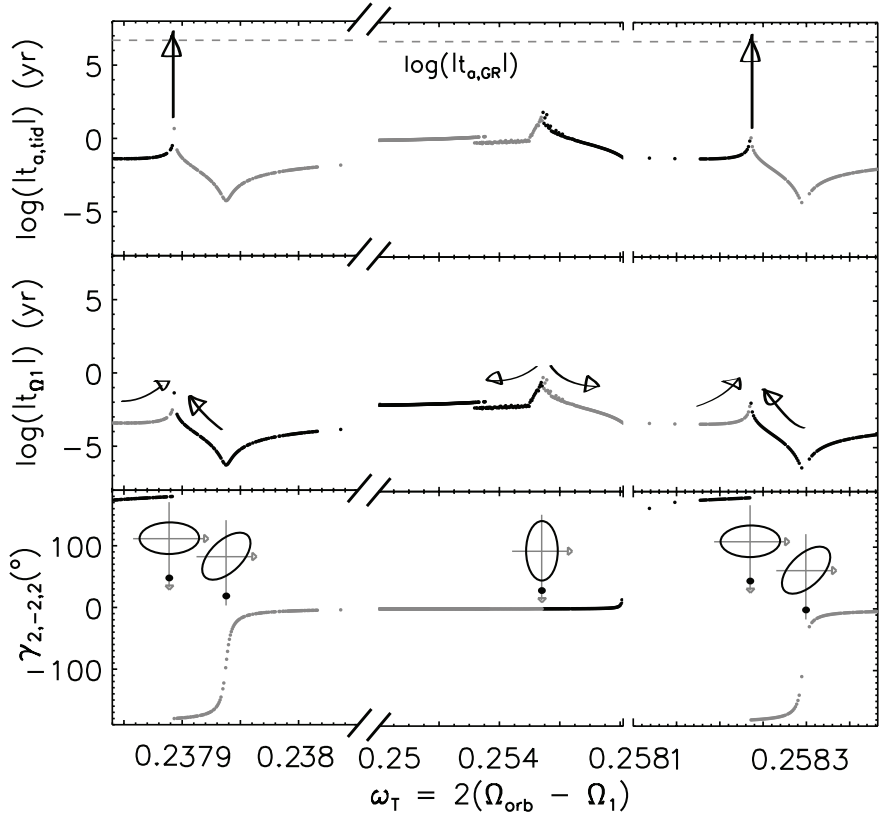}
\caption{Resonances with the modes $g_{\rm 9}$ and $g_{\rm 10}$ and
  antiresonance locking. The various symbols are as in
  Fig.~\ref{fig:tidalTimescalesLog_vsAsynch_fullRange_broken} with the
  addition of $\gamma_{2,-2,2}$ and a cartoon depicting the
  orientation of the tidal bulges in the star's frame (the
  black-filled circle denotes the companion). 
  The arrows indicate the directions towards which dynamic tides drive
  a J0651-Like binary (see text); antiresonances where   
  $\dot{\Omega}_{\rm 1}< 0$ and $\dot{\Omega}_{\rm 1} > 0$ on the left side
and right side, respectively, are trapping.  Once
  trapped, $\omega_T = 2\left(\Omega_\mathrm{orb} - \Omega_1\right)$
  is constant, while the subsequent GW emission implies
  $\dot{\Omega}_\mathrm{orb} = \dot{\Omega}_1 > 0$, so that $\Omega_1
  / \Omega_\mathrm{orb} \to 1$.  Note that the x-axis in this figure
  is not uniform.}
\label{fig:tidalTimescalesJ0651_g9g10resDetail}
\end{figure}

Fig.~\ref{fig:tidalTimescalesJ0651_g9g10resDetail} describes in detail the spin
evolution that results in antiresonance locking.
For a binary on the left of the leftmost peak in the figure,
where $\dot{\Omega}_{\rm 1}< 0$, dynamic tides act on
$\Omega_{\rm 1}$ and move the binary to the right eventually reaching
the antiresonance.  At antiresonance $t_{\rm a, tid}>
t_{\rm a, GR}$ and GR drives orbital shrinkage without affecting
$\Omega_{\rm 1}$. As $\Omega_{\rm orb}$ increases and the binary moves
to the right of the peak, dynamic tides act on the spin to bring it back at
antiresonance. This stable antiresonance locking occurs for each
antiresonance with $\dot{\Omega}_{\rm 1}< 0$ and $\dot{\Omega}_{\rm 1} > 0$ on the left side
and right side, respectively.  On the
other hand, the antiresonance in the middle panel of
Fig.~\ref{fig:tidalTimescalesJ0651_g9g10resDetail} and all regular resonances
are unstable, as dynamic tides quickly drive the binary out of them.

Mathematically, antiresonance locking is analogous to typical resonance locking (as discussed e.g., by \citealt{WitteSavonije1999}).
If the star is locked at an antiresonance frequency $\omega_{\alpha}$, 
then $\dot{\omega}_{\rm T} = \dot{\omega}_{\alpha}$. 
Assuming the background model doesn't evolve yields $\dot{\omega}_{\alpha}\,=\,0$ 
and thus $\dot{\omega}_{\rm T}\,=\,0$.
The latter assumption is justified here as for the 
MESA model $t_{\rm cool}\simeq\,90\,$Myr and it is longer than the
relevant timescales considered. As $\omega_{\rm T}$ is constant, 
$\dot{\Omega}_{\rm orb}\,=\,\dot{\Omega}_{\rm 1}$ and, since at antiresonance locking
 GR shrinks the orbit, $\dot{\Omega}_{\rm orb} = \dot{\Omega}_{\rm 1} > 0$. 
 The latter implies that the combination of orbital shrinkage due to GR
 and of antiresonance locking due to dynamic tides brings the He WD to synchronization with the
companion's orbital motion on the GR timescale. 
In fact, defining $r = \Omega_{\rm 1}/\Omega_{\rm orb}$, we find $\dot{r}\propto\dot{\Omega}_{\rm orb}(\Omega_{\rm orb}-\Omega_{\rm 1})> 0$. Since the absolute difference between 
 $\Omega_{\rm 1}$ and $\Omega_{\rm orb}$  is
fixed by the antiresonance condition and both $\left|\Omega_{\rm 1}\right|$ and $\left|\Omega_{\rm orb}\right|$
are increasing,  $r \to 1$ on a timescale 
$t_{\rm r}\,=\,r/\dot{r}\,=\,-(2/3)t_{\rm GR} \Omega_{\rm 1}(\Omega_{\rm orb}-\Omega_{\rm 1})^{-1}$.  

Consequently, the binary, while locked in antiresonance, 
it is driven to synchronism because of the GR-driven orbital decay. 
Whether J0561 has achieved synchronism already or it remains locked at antiresonance, 
still on its way to synchronism depends on the relative magnitude of the timescales involved 
($t_{\rm r}, t_{\rm GR}$, and $t_{\rm cool}$), as well as their history, 
linked to the initial properties of the system. 
Answering this question requires coupled time intergration for a range of initial conditions, 
an analysis that is beyond the scope of the present study focusing on the new 
phenomenon of antiresonance locking; such a study will be the subject of future work. 

\begin{figure} [!h]
\epsscale{0.7}
\plotone{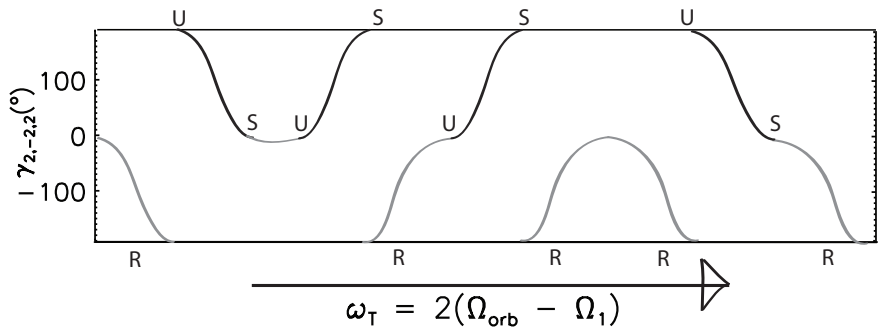}
\caption{Possible combinations for $\gamma_{2, -2, 2}$ in crossing two
  consecutive resonances (R), when the crossing occurs at $\gamma_{2,
    -2, 2}\,=\,-\pi/2$. Black and grey lines denote positive and
  negative $\gamma_{2, -2, 2}$, respectively. We use ``S'' and ``U''
  to mark stable and unstable antiresonances.}
\label{fig:schematicCombinations.eps}
\end{figure}

Even though it is difficult to identify all the stable antiresonances
in Fig.~\ref{fig:tidalTimescalesLog_vsAsynch_fullRange_broken} at the
shown resolution, the number of antiresonances expected between two
resonances can be determined from the change in the angle $\gamma_{2,
  -2, 2}$ in the passage through two consecutive resonances, as shown
in Fig.~\ref{fig:schematicCombinations.eps}. This illustration shows
all possible combinations in $\gamma_{2, -2, 2}$ when the resonance
occurs at $\gamma_{2, -2, 2}\,=\,-\pi/2$ (a similar picture can be
drawn for resonances occurring at $\gamma_{2, -2, 2}\,=\,\pi/2$).  The
number of antiresonances expected is either none, two, or four. For
the latter two cases, we expect alternating stable and unstable
antiresonances. The effect of resonances and unstable antiresonances
is to drive the binary towards the closest stable antiresonace. 
This behavior is generic for any object under the influence of
non-adiabadic dynamic tides, if the evolution of its properties are neglected; 
the locations of the resonances and
antiresonances depend on the details of the object's structure, but
the general picture of antiresonance locking depends only on the
continuity of the phase $\gamma_{l,m,k}$ between resonances.

\section{Conclusions} \label{sec:Conclusions}
In this Letter, we show that dissipative dynamic tides potentially
play a key role in explaining the recently measured orbital decay of J0651 \citep{HermesEtAl2012}, 
even though their effects are not directly apparent.
In fact, in the most probable scenario in which the binary was asynchronous at birth, 
the agreement between the observed orbital decay and the GR prediction 
is achieved through the new phenomenon of antiresonance locking identified here.
Assuming the He WD doesn't evolve on a relevant timescale, dynamic tides naturally lock the system at stable 
antiresonances with the star's eigenfrequencies. While locked, GR drives the evolution of the orbital separation, 
with dynamic tides
maintaining the antiresonance condition and drive the binary to synchronism \emph{on the GR timescale}.  
Time integration of the binary orbit for a wide range of initial conditions and a self-consistent calculation of all the relevant timescales
will allow to assess whether the system is still locked and away from synchronism or if it has had 
time to reach synchronism. 
 Future measurements of the He WD spin can potentially provide a check of the theoretical predictions presented here.

 \acknowledgments


\begin{acknowledgements}
 We are grateful to C. Deloye, A. Barker, Y. Lithwick, M. Kilic,
 A. Gianninas, and W. Brown for useful discussions during the
 development of this project. Simulations were performed on the
 computing cluster {\tt Fugu} available to the Theoretical
 Astrophysics group at Northwestern and partially funded by NSF grant
 PHY--0619274 to VK. This work was supported by NASA Award NNX09AJ56G
 to V.K.
 
\end{acknowledgements}
\newpage

\end{document}